\documentclass{ws-p8-50x6-00}

\def\PRC{{\em Phys. Rev.} C}
\def\EPJC{{\em Eur. Phys. J.} C}

\begin{document}

\title{
Examining CP Symmetry in Strange Baryon Decays
}

\author{K. B. Luk}

\address{
Department of Physics, University of California and
Physics Division, Lawrence Berkeley National Laboratory,
Berkeley CA 94720, USA\\
E-mail: k\_luk@lbl.gov
}
\maketitle
\begin{center}
representing the Fermilab E756 and HyperCP Collaborations
\end{center}
\vspace{5pt}
\small{
R. A. Burnstein,$^{6,b}$
A. Chakravorty,$^{6,b}$
A. Chan,$^{3,b}$
Y. C. Chen,$^{3,b}$
W. S. Choong,$^{1,2,b}$
K. Clark,$^{11,b}$
T. Diehl,$^{4,a}$
E. C. Dukes,$^{12,b}$
C. Durandet,$^{12,b}$
J. Duryea,$^{9,a}$
J. Felix,$^{5,b}$
G. Gidal,$^{1,b}$
G. Guglielmo,$^{4,a}$
H. R. Gustafson,$^{8,b}$
K. Heller,$^{9,a}$
C. Ho,$^{3,b}$
P. M. Ho,$^{1,a}$
T. Holmstrom,$^{12,b}$
M. Huang,$^{12,b}$
C. James,$^{4,a,b}$
M. Jenkins,$^{11,b}$
K. Johns,$^{9,a}$
T. Jones,$^{1,b}$
D. M. Kaplan,$^{6,b}$
L. M. Lederman,$^{6,b}$
N. Leros,$^{7,b}$
F. Lopez,$^{8,b}$
M. J. Longo,$^{8,a,b}$
L. Lu,$^{12,b}$
W. Luebke,$^{6,b}$
K. Nelson,$^{12,b}$
H. K. Park,$^{8,b}$
J. P. Perroud,$^{7,b}$
D. Rajaram,$^{6,b}$
R. Rameika,$^{4,a}$
H. A. Rubin,$^{6,b}$
S. Teige,$^{6,a}$
P. K. Teng,$^{3,b}$
G. Thomson,$^{10,a}$ 
J. Volks,$^{4,b}$
C. G. White,$^{6,b}$
S. L. White,$^{6,b}$
P. Zyla$^{1,b}$\\
\begin{quote}
$^a$ Fermilab E756 Collaboration \\
$^b$ Fermilab HyperCP Collaboration \\
\it{
$^1$ Lawrence Berkeley National Laboratory, Berkeley, California 94720\\
$^2$ University of California, Berkeley, California 94720\\
$^3$ Academia Sinica, Nankang, Taipei 11529, Taiwan, R.O.C.\\
$^4$ Fermilab, Batavia, Illinois 60510\\
$^5$ University of Guanajuato, 37000 Leon, Mexico\\
$^6$ Illinois Institute of Technology, Chicago, IL 60616\\
$^7$ University of Lausanne, CH-1015 Lausanne, Switzerland\\
$^8$ University of Michigan, Ann Arbor, Michigan 48109\\
$^9$ University of Minnesota, Minneapolis, Minnesota 55455\\
$^{10}$ Rutgers--The State University, Piscataway, New Jersey 08854\\
$^{11}$ University of South Alabama, Mobile, Alabama 36688\\
$^{12}$ University of Virginia, Charlottesville, Virginia 22901
}
\end{quote}
}
\vspace{10pt}

\abstracts{
Non-conservation of CP symmetry can manisfest itself in non-leptonic 
hyperon decays as a difference in the decay parameter between the 
strange-baryon decay and its charge conjugate.
By comparing the decay distribution in the $\Lambda$ helicity frame 
for the decay sequence $\Xi^{-} \to \Lambda \pi^{-}$, 
$\Lambda \to p \pi^{-}$ 
with that of $\overline{\Xi}^{+}$ decay, E756 at Fermilab did not observe any 
CP-odd effect at the $10^{-2}$ level.
The status of a follow-up experiment, HyperCP (FNAL E871), to search 
for CP violation in charged $\Xi-\Lambda$ decay with a sensitivity of $10^{-4}$ 
is also presented.
}

\section{Introduction}
The Standard Model, as well as many other models, predicts the existence of CP
violation in non-leptonic strange-baryon decays.\cite{Pakvasa1}
One approach to observe this CP-odd effect is to compare the decay 
distribution of a strange baryon with that of its charge conjugate. 
In the non-leptonic weak decay of hyperon, parity violation leads to a 
forward-backward asymmetric distribution of the daughter particles with 
respect to the spin of the hyperon:
\vspace{5pt}
\begin{equation} \label{eq:decay}
\frac{dn}{dcos\theta} = \frac{1}{2} \big( 1 + \alpha {\mathbf{P}}
\cdot \hat{\mathbf{p}} \big)
=  \frac{1}{2} \big( 1 + \alpha P cos \theta \big)
\vspace{5pt}
\end{equation}
where $\alpha$ is a parameter quantifying the degree of parity violation 
in the decay, 
$\mathbf{P}$ is the polarization of the hyperon, 
and $\hat{\mathbf{p}}$ is the momentum unit vector of the 
daughter baryon in the rest frame of the hyperon.

Under CP transformation, the strange-baryon decay is related to the 
corresponding decay of its anti-particle with the condition that 
\vspace{5pt}
\begin{equation} \label{eq:alfa}
\overline{\alpha} = -\alpha~.
\vspace{5pt}
\end{equation}
Thus, any violation of equation (\ref{eq:alfa}) would signal the breaking of CP 
symmetry in the decay.
Alternatively we can define a parameter
\vspace{5pt}
\begin{equation} \label{eq:asy}
A = \frac {\alpha+\overline{\alpha}}{\alpha-\overline{\alpha}}
\vspace{5pt}
\end{equation}
for measuring the amount of CP violation in the decay.
The latest predictions of $A$ for hyperon decays 
are summarized in reference 1.

To date all searches of CP nonconservation in strange-baryon sector  
focussed on the $\Lambda \to p \pi$ decay, with $\Lambda$'s and 
$\overline{\Lambda}$'s produced in $p\overline{p}$ collisions~\cite{ISR,PS185}
or from the $J/{\Psi}$ decays.\cite{dm2}
The best result of $0.013 \pm 0.022$ for $A_{\Lambda}$ came from 
LEAR PS185.\cite{PS185}

In this talk, a new approach for studying CP symmetry in strange-baryon 
decay is presented.
According to equation~(\ref{eq:decay}) $\Lambda$ hyperons with precisely known 
polarization are needed for determining $\alpha_{\Lambda}$.
Significantly polarized $\Lambda$'s can be created from charged 
$\Xi \to \Lambda \pi$ decays.
In this case, the polarization of the $\Lambda$ is given by
\vspace{5pt}
\begin{equation} \label{eq:lampol}
{\mathbf{P}}_{\Lambda} = \frac{(\al_{\Xi} + {\mathbf{P}}_{\Xi} 
\cdot \hat{\mathbf{\Lambda}})
\hat{\mathbf{\Lambda}}
+ \beta_{\Xi} {\mathbf{P}}_{\Xi} \times \hat{\mathbf{\Lambda}}
+ \gamma_{\Xi} \hat{\mathbf{\Lambda}} \times ({\mathbf{P}}_{\Xi} \times
\hat{\mathbf{\Lambda}})}
{\big( 1 + \al_{\Xi} {\mathbf{P}}_{\Xi} \cdot \hat{\mathbf{\Lambda}} \big)}
\vspace{5pt}
\end{equation}
where $\hat{\mathbf{\Lambda}}$ is the unit vector along the $\Lambda$ 
momentum in the $\Xi$ rest frame, $\beta_{\Xi}$ and $\gamma_{\Xi}$ are the 
other two decay parameters of the $\Xi \to \Lambda \pi$ decay, and 
$\mathbf{P}_{\Xi}$ is the polarization of the $\Xi$.
In the helicity frame of the $\Lambda$ the angular distribution of the proton 
is
\vspace{5pt}
\begin{equation} \label{eq:helicity}
\frac{dn}{dcos\theta_{p\Lambda}} 
= \frac{1}{2} \big( 1 + \alpha_{\Lambda} \alpha_{\Xi}~ 
cos \theta_{p\Lambda} \big)~,
\vspace{5pt}
\end{equation}
independent of the polarization of the $\Xi$.
A new parameter, $A_{\Xi \Lambda}$, defined as
\vspace{5pt}
\begin{equation} \label{eq:asy2}
A_{\Xi\Lambda} = \frac {\alpha_{\Xi} \alpha_{\Lambda} -
\alpha_{\overline{\Xi}} \alpha_{\overline{\Lambda}}}
{\alpha_{\Xi} \alpha_{\Lambda} +
\alpha_{\overline{\Xi}} \alpha_{\overline{\Lambda}}}
\approx A_{\Xi} + A_{\Lambda}
\vspace{5pt}
\end{equation}
is used for searching for CP asymmetry in the $\Xi-\Lambda$ decay 
sequence. 
This new scheme was first demonstrated by E756 at Fermilab with polarized
$\Xi$. 
A value of $0.012 \pm 0.014$ for $A_{\Xi \Lambda}$ was obtained.\cite{e756cp}
Motivated by this successful study, a dedicated experiment, 
HyperCP (E871), was proposed and is currently taking data at Fermilab to 
reach a projected sensitivity of $10^{-4}$.
The remainder of this talk will highlight the analysis of E756 and present 
the status of HyperCP. 

\section{FNAL E756}\label{sec:e756}
The primary goals of FNAL E756 were to measure the magnetic moment of the 
$\Omega^{-}$~\cite{diehl} and 
to study the production polarization of hyperons.\cite{duryea,PMH,luk}
Along with the $\Omega^{-}$'s a large sample of $\Xi^{-}$ decays was obtained.

This experiment was performed in the FNAL Proton-Center beam line during the
1987 fixed-target run.
The plan view of the E756 spectrometer is shown in 
Fig.1.
Polarized $\Xi^{-}$ hyperons were produced with an 800-GeV proton beam 
striking a 2~mm $\times$ 2~mm $\times$ 9.2~cm-long 
beryllium target at an angle of 2.4 mrad in the vertical plane.
After the strange baryons emerged from a curved collimator in a 7.3~m-long 
hyperon magnet, with its field in the vertical direction, 
\begin{figure} [htb]
\label{fig:E756spectrometer}
\epsfxsize=24.5pc
\epsfbox{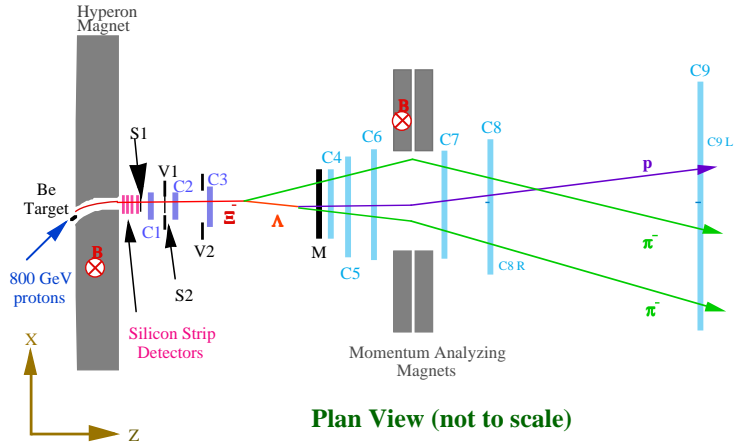}
\caption{Plan view of E756 spectrometer (not to scale).}
\end{figure}
they were allowed to decay in a 26~m-long region. 
The charged particles in the decay sequence $\Xi^{-} \to \Lambda \pi^{-}$ and
$\Lambda \to p \pi^{-}$ were momentum analyzed with a simple spectrometer
consisting of scintillation counters, silicon strip detectors, multiwire 
proportional chambers, and two dipole magnets.
The trigger was designed to look for events with at least two oppositely 
charged tracks downstream of the analysis magnets and with three or  
four tracks upstream.
The targetting angle was cycled between 2.4 mrad and -2.4 mrad every few 
hours to minimize temporal variations between runs.
For each production angle, roughly equal number of events were also 
collected with the polarity of the analysis magnets reversed 
and the left-right requirements of C8 and C9 interchanged.

Ten days were spent to collect $\overline{\Xi}^{+}$ decays, 
the incident proton intensity was reduced from $3 \times 10^{10}$ 
per 20-s spill 
to $1 \times 10^{10}$ per spill, to maintain a similar singles rate in 
the spectrometer as the $\Xi^-$ run.
The polarity of the hyperon magnet was also reversed but the  
trigger requirements and the production angles were unchanged, thus 
reducing potential biases between 
the $\Xi^{-}$ and $\overline{\Xi}^{+}$ runs.

For the CP-symmetry study, 
the hyperon magnetic field was set for selecting $\Xi$'s with 
monenta between 240 GeV/c and 450 GeV/c. 
 
Both the $\Xi^{-}$ and $\overline{\Xi}^{+}$ data were processed with the same 
reconstruction program and were subjected to identical event-selection 
requirements.
Further details of the experiment and the event selection can be found in 
reference 8.

\begin{figure} [htb]
\label{fig:hmc}
\epsfxsize=24pc
\epsfbox{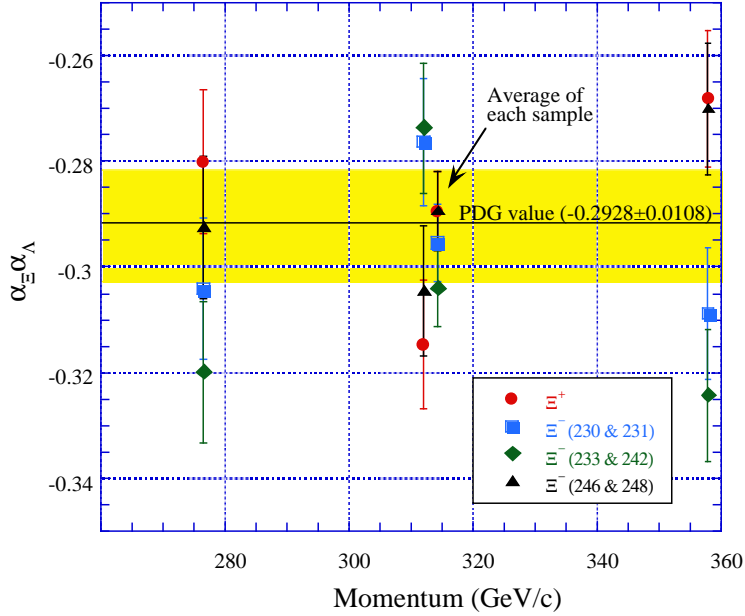}
\caption{E756 results on $\alpha_{\Xi}\alpha_{\Lambda}$ as a function of 
the momentum of the $\Xi$. The shaded area is an one-standard-deviation band 
centered at the world average.}
\end{figure}

Two different techniques were employed to extract $A_{\Xi\Lambda}$. 
In the first approach, the value of $\alpha_{\Xi}\alpha_{\Lambda}$ was 
determined with the Hybrid Monte Carlo method\cite{PMH} 
for the $\Xi^{-}$ and the $\overline{\Xi}^{+}$ samples separately.
Based on about 70,000 $\overline{\Xi}^{+}$ decays, $\alpha_{\overline{\Xi}}
\alpha_{\overline{\Lambda}}$ was found to be $-0.2894 \pm 0.0073$. 
From three independent $\Xi^{-}$ samples, each with approximately 
70,000 events, $\alpha_{\Xi}\alpha_{\Lambda}$ was 
$-0.2955 \pm 0.0073$, $-0.3041 \pm 0.0073$, and $-0.2894 \pm 0.0073$, 
giving a mean of $-0.2963 \pm 0.0042$.
As shown in Fig.~\ref{fig:hmc} these results were in good agreement 
with the world 
average\cite{PDG} and were stable with respect to the momentum of $\Xi$.
This method gave a value of $0.012 \pm 0.014$ for $A_{\Xi\Lambda}$, 
with insignificant 
systematic error that was estimated by varying event-selection 
requirements.

In the second approach, the difference in $\alpha_{\Xi}\alpha_{\Lambda}$ 
between $\Xi^-$ and $\overline{\Xi}^+$ was 
determined directly without unfolding the acceptance in cos$\theta_{p\Lambda}$.
For two data samples, a comparison of the cos$\theta_{p\Lambda}$ 
distributions can be defined by
\vspace{5pt}
\begin{equation} \label{eq:ratio}
R(cos \theta_{p\Lambda}) = 
\frac{\epsilon_1(cos \theta_{p\Lambda})} 
{\epsilon_2(cos \theta_{p\Lambda})}
\frac{[ 1 + (\alpha_{\Lambda} \alpha_{\Xi})_1 cos \theta_{p\Lambda}]}
{[ 1 + (\alpha_{\Lambda} \alpha_{\Xi})_2 cos \theta_{p\Lambda}]}
\vspace{5pt}
\end{equation}
where $R(cos \theta_{p\Lambda})$ is the ratio of the probabilities of  
getting cos$\theta_{p\Lambda}$ in the two samples, 
and the $\epsilon$'s are the acceptance functions of the cos$\theta_{p\Lambda}$ 
distributions.

When two sets of $\Xi^-$ events are compared, $R$ is a measure of how 
well the acceptances agree.
Without any corrections, 
detailed studies showed that the acceptance in cos$\theta_{p\Lambda}$ 
was strongly dependent 
on the momentum of the $\Xi^-$, but was insensitive to the polarization of the
$\Xi^-$ or other variations in the experiment down to a few $\times 10^{-3}$ 
level.\cite{e871}
This unique feature is due to the fact that the $\hat{\mathbf{\Lambda}}$ 
defining the helicity frame  
changes from event to event over the entire phase space in the $\Xi$ rest frame.
Any systematic bias due to imperfection of the experiment in the laboratory 
is mapped into a wide range of cos$\theta_{p\Lambda}$, thus highly diluted.

In the study of CP symmetry, a sample of $\Xi^-$ events was selected in such 
a way that 
the resulting $\Xi^-$ momentum spectrum was identical to that of the 
$\overline{\Xi}^+$ sample.
This removed the difference due to different mechanism for producing   
particles and anti-particles by protons, and 
ensured that $\epsilon(cos \theta_{p\Lambda})$ was identical for 
both data sets.
With $\alpha_{\Xi}\alpha_{\Lambda}$ taken to be -0.2928\cite{PDG} 
the difference in $\alpha_{\Xi}\alpha_{\Lambda}$ between the $\Xi^-$ and 
$\overline{\Xi}^+$ samples was determined by fitting $R$ as a function of 
cos$\theta_{p\Lambda}$ according to equation~(\ref{eq:ratio}).
With approximately 70,000 $\Xi^-$ events along with equal number of 
$\overline{\Xi}^+$ decays, the difference was $-0.011 \pm 0.009$.
This implied that $A_{\Xi\Lambda}$ was $0.019 \pm 0.015$, which was 
consistent with the result obtained with the Hybrid Monte Carlo method.
As a check, another sample of $\Xi^-$ events was picked to repeat 
the measurement which yielded 
a result of $0.008 \pm 0.015$ for $A_{\Xi\Lambda}$.

\section{HyperCP (FNAL E871)}\label{sec:e871}
HyperCP is a dedicated experiment designed to search for CP violation 
in charged-$\Xi$-$\Lambda$ decay with about 100 times better sensitivity 
than E756.
This requires a high-rate spectrometer which is sufficiently simple for 
controlling systematic effects.

\begin{figure} [htb]
\label{fig:e871spectrometer}
\epsfxsize=25pc
\epsfbox{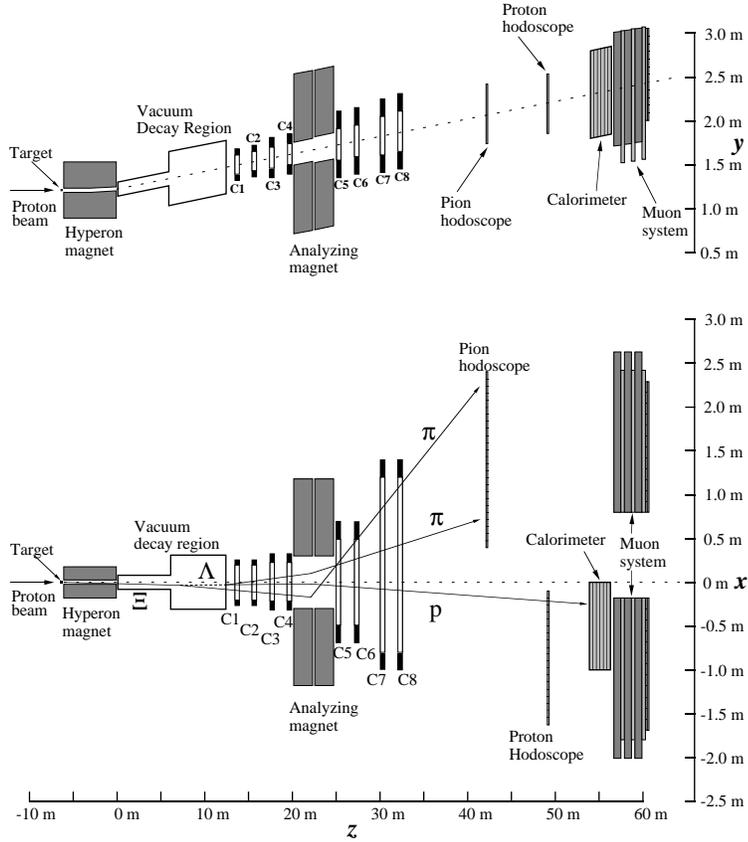}
\vspace{20pt}
\caption{Spectrometer of HyperCP.}
\end{figure}
As shown in Fig.~\ref{fig:e871spectrometer}, 
an 800-GeV proton beam with a typical intensity of  
$ 7.5 \times 10^{9}$ per sec incidents on a 2~mm $\times$ 2~mm $\times$ 
3~cm-long copper target at 0~mrad.
The secondary beam is bent through a 6.1~m-long shaped channel in the 
vertical plane by a dipole magnet with a field of about 1.667~T. 
The momentum range of the selected $\Xi^-$'s is between 120 GeV/c and 220 GeV/c.
The acceptance of this channel is about a factor of two larger than the one 
in E756.
Downstream of a 13~m-long evacuated decay region is a spectrometer 
consisting of eight multiwire proportional chambers and two analysis magnets.
There are about two times more anode planes and finer wire pitch than the E756 
spectrometer, offering better reconstruction efficiency and better 
momentum resolution.\footnote{The muon system at the end of the spectrometer 
is used for studying rare decays.}
The trigger elements are a hadron calorimeter for detecting
particles with energy greater than 70 GeV at close to 100\% efficiency, 
and two hodoscopes 
for finding opposite-sign charged tracks downstream of the analysis magnets.

Building upon the technique developed in E756, 
$\overline{\Xi}^+$ events are collected by switching to an approximately 
2~cm-long copper target for sustaining a singles rate comparable to that 
for the $\Xi^-$ run, and the 
polarities of the hyperon magnet and the spectrometer magnets are reversed 
while keeping the trigger unchanged.
The $\Xi^-$ and $\overline{\Xi}^+$ modes are rotated every several hours 
to ensure adjacent runs having similar run conditions.

The experiment had a successful first run in the 1997 fixed-target program, 
collecting about 30 billion $\Xi$ triggers on tape.\cite{white}  
Due to an increase of the machine duty cycle, from about 40\% to 50\%, 
the DAQ had to be upgraded for the 1999 run.
This was realized by increased buffering, reduced event size through 
hardware data compression, improved software, better layout of 
the DAQ architecture, and the use of the faster Exabyte 8705 tape drives.  
A typical DAQ bandwidth of 23 Mbyte/s, about a factor of two improvement 
over the 1997 setup, and a live time of better than 75\% is achieved.

Other improvements to the spectrometer, all of which 
resulted in better quality data, are also implemented in the 1999 run. 
These include a detailed study of the relative alignment of elements 
in the target area by beam scanning to ensure that the targetting  
angle is close to 0 mrad and the target is centered at the entrance of the 
collimator, installing a new proton  
trigger hodoscope with a 
second set of counters to improve efficiency determination, 
installing a segmented beam monitor for handling the intense secondary beam, 
and running the downstream chambers with 
CF$_{4}$-isobutane gas mixture to further reduce background hits. 

Fig. 4 is a comparison of the 1997 and the 1999 data.
Due to an adjustment in the target position the $\Xi$ momentum in the 1999 run 
is lowered, closer to the Monte Carlo prediction. 
The resolution of the $\Lambda\pi$ invariant mass of the two runs is 
about 1.6~MeV/c$^2$, 
which is about 0.8 MeV/c$^2$ better than in E756.
\begin{figure} [b]
\label{fig:97-99-comparison}
\epsfxsize=25pc
\epsfbox{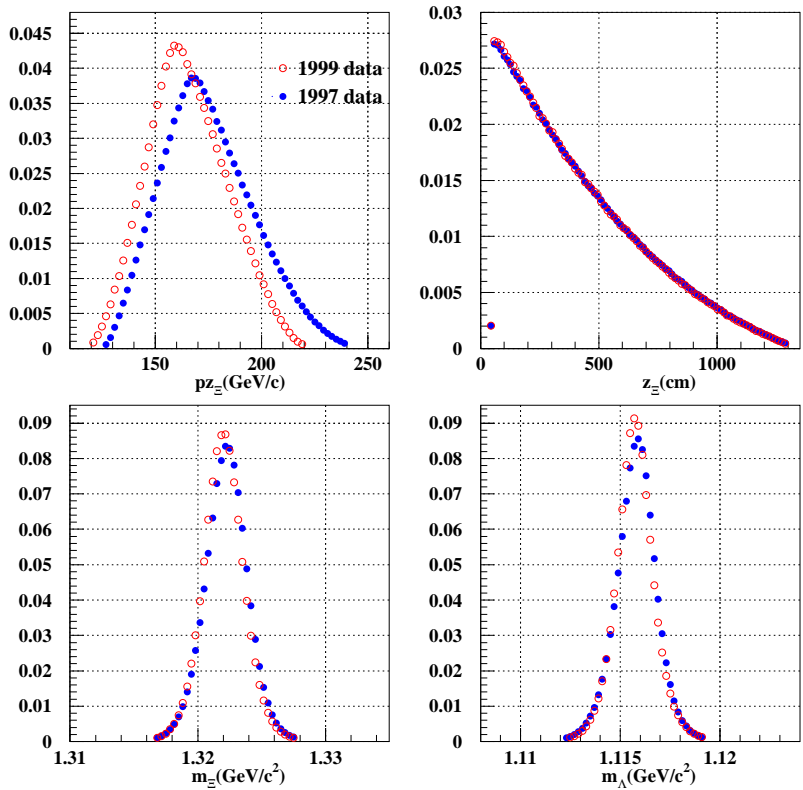}
\vspace{20pt}
\caption{Comparison of momentum, decay position, $\Lambda\pi$ invariant mass, 
and p$\pi$ invariant mass between the 1997 and the 1999 run in HyperCP.}
\end{figure}

The 1999 run is ongoing.
By the end of the run, around the middle of January, 2000, 
about a factor of two more data than the 1997 run is expected to be logged.
The combination of the two runs is expected to 
yield about $3 \times 10^9$ reconstructed 
$\Xi^-$ decays and $0.75 \times 10^9$ $\overline{\Xi}^+$ events.
This corresponds to a projected statistical 
sensitivity for the strange-baryon CP asymmetry of about $1.3 \times 10^{-4}$, 
which is comparable to the proposed goal. 

\section{Conclusion}
Based on approximately 70,000 $\overline{\Xi}^+$ and 210,000 $\Xi^-$ decays 
the Fermilab E756 Collaboration has searched for CP violation 
in hyperon decay by measuring $A_{\Xi\Lambda}$. 
A result of $0.012 \pm 0.014$ was obtained. 
Systematic biases to the measurement, even without making any 
correction, have been shown to be insignificant  
down to the $10^{-3}$ level.

HyperCP has two successful runs, amassing probably the largest sample 
of strange-baryon decays in the world. 
By the end of the 1999 run, the projected numbers of fully reconstructed 
$\Xi^-$ and $\overline{\Xi}^+$ events are approximately $3 \times 10^{9}$ and 
$0.75 \times 10^9$ respectively.
This should yield a statistical precision of $1.3 \times 10^{-4}$ for 
$A_{\Xi\Lambda}$.

\section*{Acknowledgments}
K. B. Luk would like to thank the organizers, Professor George Hou 
and Dr. Hai-Yang Cheng, for the invitation and their hospitality. 
My gratitude also goes to Sandip Pakvasa, German Valencia, and X. G. He 
for many stimulating discussions.
This work was supported by the Director, Office of Energy Research, Office of
High Energy and Nuclear Physics, Division of High Energy Physics of the U.S.
Department of Energy under Contract DE-AC03-76SF00098, the National Science 
Foundation, and the National Science Council of Taiwan, R.O.C..

\end{document}